# Research on OPF control of three-phase four-wire low-voltage distribution network considering uncertainty

Rui Wang, Xiaoqing Bai*, Shengquan Huang, and Shoupu Wei

Guangxi Key Laboratory of Power System Optimization and Energy-Saving Technology, College of Electrical Engineering, Guangxi University, Nanning 530004, China.

2112392095@st.gxu.edu.cn ,* baixq@gxu.edu.cn , 2212392044@st.gxu.edu.cn , 2112392098@st.gxu.edu.cn
*: Correspondent author: Dr. Xiaoqing Bai

**Abstract**

As power systems become more complex and uncertain, low-voltage distribution networks face numerous challenges, including three-phase imbalances caused by asymmetrical loads and distributed energy resources. We propose a robust stochastic optimization (RSO) based optimal power flow (OPF) control method for three-phase, four-wire low-voltage distribution networks that consider uncertainty to address these issues. Using historical data and deep learning classification methods, the proposed method simulates optimal system behaviour without requiring communication infrastructure. The simulation results verify that the proposed method effectively controls the voltage and current amplitude while minimizing the operational cost and three-phase imbalance within acceptable limits. The proposed method shows promise for managing uncertainties and optimizing performance in low-voltage distribution networks.

## 1. Introduction

In the past few years, the economy has grown significantly, but as a result, there has been an increase in worries about environmental contamination and the shortage of energy resources. As a response, renewable energy generation, such as photovoltaic, has been vigorously promoted [1-4]. To minimize the potential environmental hazards that result from the indiscriminate consumption of fossil fuels, different governments and organizations have been urging customers to utilize renewable energy alternatives such as rooftop PV. Nonetheless, studies have indicated that installing rooftop PV may affect the node voltage and network loss indicators of the original distribution network, leading to overvoltage issues during the day [5]. Additionally, the three-phase four-wire structure of a low-voltage distribution network may produce an unbalanced phenomenon during times of asymmetric loads and line parameters, resulting in increased line losses and zero sequence current. Therefore, it is essential to study the overvoltage problem caused by uncertainties under such a structure.

The purpose of distribution network operation control is to ensure the reliability of power supply and economical operation, which essentially belongs to the category of OPF [6]. Since the three-phase four-wire low-voltage distribution network is more complex in terms of model and algorithm, the current research focuses on the three-phase, medium-voltage distribution network. Several researchers have developed the OPF model either as a linear or non-linear programming model, which they have then directly solved using artificial intelligence algorithms (e.g., genetic algorithm [7], particle swarm algorithm [8], etc.), but there are disadvantages such as slow solution speed and easy to fall into local optimum; some other scholars have convexly relaxed the established nonlinear model by first eliminating the phase angles of voltage and current, and then convexified the original model using second-order cone relaxation[9-11]. However, the elimination of the phase angle results in the inability to calculate the voltage and current on the neutral line, which makes the method unsuitable for usage in three-phase four-wire lines. Reference [12] proposes a multiphase OPF algorithm suitable for unbalanced distribution systems, which can be applied to enhance the performance of smart grid systems. On the other hand, reference [13] introduces an OPF formulation for four-wire networks, addressing robustness and scalability issues. These references provided valuable insights into developing efficient power flow algorithms in different network environments. These have also informed our research and contributed to developing our proposed three-phase four-wire OPF algorithm designed explicitly for low-voltage distribution networks with high renewable energy sources penetration levels.

In the management of distributed energy resources (DERs), the available operating solutions for control can be broadly categorized as centralized, distributed, and decentralized, depending on the communication infrastructure that is in place. Centralized solutions typically necessitate a comprehensive monitoring and communication infrastructure and utilize high-performance control technologies based on optimization techniques. For instance, in reference [14], a layered coordinated voltage correction scheme is proposed that is suitable for active distribution networks. Similarly, reference [15] proposes a strategy that allows optimization of the hardware required for implementation and has good dynamic performance. Reference [16] proposes a centralized approach with an existing communication infrastructure that can operate in islanded mode when necessary. While a control and monitoring strategy for a 0.4kV distribution network is proposed in reference [17], the issue of integrating renewable energy sources into the power grid remains unaddressed, posing a significant challenge. On the other hand, reference [18] introduces a novel strategy for load reduction that is based on the DC voltage level, which is applied to DC distribution networks.

Decentralized control policies are commonly employed in Distribution Networks (DNs) and have been incorporated into multiple grid codes. One of the advantages of utilizing such policies is their ease of implementation and relatively low cost. Decentralized control policies do not require a communication infrastructure, which, in turn, minimizes the necessary investment for implementation [19]. However, such methods commonly adopt a "one size fits all" approach, utilizing identical control parameters across all DNs, various generator types, and varying operating conditions [20]. This approach takes the same control strategy when generating failures caused by various factors, which result in unanticipated complications, particularly in a rapidly evolving environment.

Distributed approaches coordinate different DERs with limited communication to achieve near-optimal operations. Although efforts have been made to narrow the gap between local and centralized approaches, these methods still necessitate certain communication infrastructure. Reference [21] proposes a two-phase multi-objective control method for distribution networks and verified the effectiveness through arithmetic examples. Reference [22] analyses an unbalanced four-wire distribution network optimization operation methodology that includes DG, which is validated and tested in a 13-node system.

To solve the problems of voltage overrun and network loss in a three-phase four-wire low-voltage distribution network, we propose an OPF control method that does not require the deployment of communication and monitoring infrastructure. Figure 1 illustrates the flow chart of the proposed method.

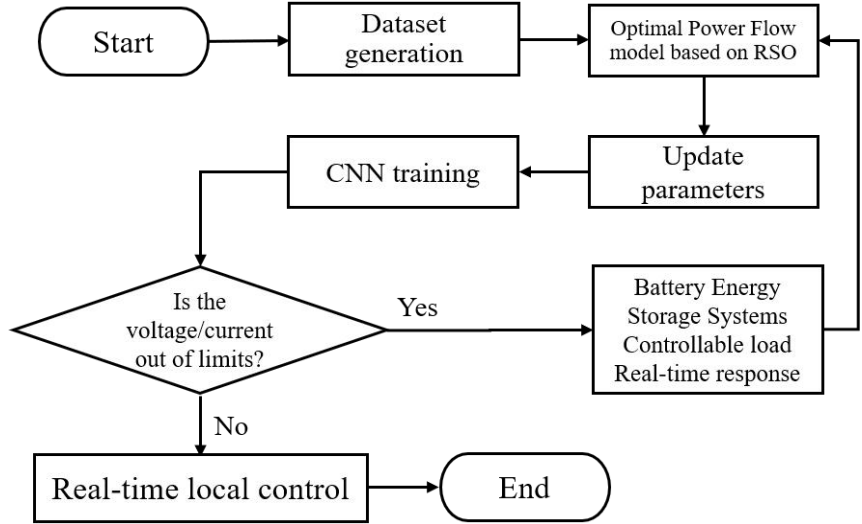

Figure 1: Flow chart of the proposed method.

In addition, local control is valuable even where communication facilities are well established, as more and more distributed resources can be efficiently utilized. More specifically, the contributions of this paper are summarized as follows:

(i) Considering RES uncertainty and unbalanced operation, an optimal power flow model (RSO-AC-OPF) is proposed based on a low-voltage distribution network and robust stochastic optimization (RSO) method.
(ii) A control design method can effectively achieve real-time optimal distribution network control, minimizing control costs and improving reliability.

The rest of the paper is organized as follows: Section 2 presents the network equations and mathematical model considering uncertainty, including the objective function and constraints. Section 3 derives the power system control method to optimize distributed energy sources. In Section 4, we perform case tests and comparative simulations. Lastly, Section 5 provides a conclusion for this paper.

## 2. Model
### 2.1. AC-OPF for the Three-phase Four-wire Network

This section presents the mathematical equations that describe the network and the models used in solving the OPF control problem for the three-phase four-wire low-voltage distribution network that is proposed in this paper.

The objective function that has been chosen includes minimizing both the control cost of DER, network losses and the expected adjustment in the worst case, just as follows:

$$S = \min(S_1 + S_2) \tag{1}$$

$$S_1 = \sum_{t=1}^{T}\left[\sum_{z\in(a,b,c)}\sum_{j=1}^{N_{bus}}(C_P \cdot P_{redu,j,z,t} + C_Q \cdot Q_{cont,j,z,t}) + \sum_{z\in(a,b,c)}\sum_{i=1}^{N_{bran}} C_p \cdot P_{loss,i,z,t}\right]\cdot \Delta t \tag{2}$$

$$S_2 = \max_{P\in\Phi}\sum_{n=1}^{N} P_n \min \sum_{t=1}^{T}[C_{PV}(P_{PV,t} - P_{PV,t}^{act})]\cdot \Delta t \tag{3}$$

$$P_{redu,j,z,t} = P_{g,j,z,t}^{\max} - P_{g,j,z,t} \tag{4}$$

$$Q_{cont,j,z,t} = |Q_{g,j,z,t}| \tag{5}$$

where $N_{bus}$ is the node number, $N_{bran}$ is the branch number; $C_P$, $C_Q$ and $C_{PV}$ are the active and reactive DG (distributed generation) costs and PV adjustment costs, respectively. $P_{g,j,z,t}$ is active power injection in the distributed generation system. $Q_{g,j,z,t}$ is reactive power injection/absorption of the distributed generation system. $P_{loss,i,z,t}$ represents the network loss on phase z of branch i at time t. $P_{PV,t}$ and $P_{PV,t}^{act}$ are the predicted and actual output of the PV system, respectively. $P_n$ denotes the probability of each discrete scenario n.

For the voltage amplitude of each node on three phases, there should be a maximum limit value according to the national standard in order to guarantee the safe operation of the power grid:

$$V_{\min} \leq |V_{j,z,t}| \leq V_{\max} \tag{6}$$

where $|V_{j,z,t}|$ is the voltage amplitude of node j at time z phase t; $V_{\min}$ and $V_{\max}$ are the minimum and maximum allowable voltage values.

The current constraint is described as

$$|I_{bran,i,z,t}| \leq I_{i,z}^{\max} \tag{7}$$

where $I_{bran,i,z,t}$ is the current in phase z of branch i at time t; $I_{i,z}^{\max}$ is the maximum allowable current of branch i.

The neutral line voltage constraint can be expressed as

$$|V_{j,neut,t}| \leq V_{j.neut.t}^{\max} \tag{8}$$

where $|V_{j,neut,t}|$ is the voltage amplitude of node j at time t on the neutral line; the maximum allowable value for the voltage of the neutral line is $V_{j.neut.t}^{\max}$.

The power balance constraints are expressed as

$$P_{in,j,z,t} = P_{g,j,z,t} - P_{d,j,z,t} - P_{BESS,j,z,t}^{char} + P_{BESS,j,z,t}^{disc} \tag{9}$$

$$Q_{in,j,z,t} = Q_{g,j,z,t} - P_{d,j,z,t} \tan \varphi + Q_{BESS,j,z,t} \tag{10}$$

where $P_{in,j,z,t}$ and $Q_{in,j,z,t}$ are the active and reactive power injected by the bus. $P_{d,j,z,t}$ and $P_{d,j,z,t} \tan \varphi$ are the requirements of active and reactive nodes after control, respectively. $\cos \varphi$ is the load power factor. BESS is short for battery energy storage system. $P_{BESS,j,z,t}^{char}$ and $P_{BESS,j,z,t}^{disc}$ represent the active power charged and discharged by the BESS, respectively.

The constraints of power flow are described as

$$I_{in,j,z,t} = [\frac{(P_{in,j,z,t} + jQ_{in,j,z,t})^*}{\bar{V}_{j,z,t}^*}] \tag{11}$$

$$I_{bran,t} = A \cdot I_{in,j,z,t} \tag{12}$$

$$\Delta V_t = B \cdot I_{bran,t} \tag{13}$$

$$V_t = V_{slac} - \Delta V_{OLTC} \cdot \rho_t + \Delta V_t \tag{14}$$

$$\rho_{min} \leq \rho_t \leq \rho_{max} \tag{15}$$

where $I_{in,j,z,t}$ represents the current injected by the Z-phase bus, $\bar{V}_{j,z,t}^*$ represents the voltage of node j phase z. $I_{bran,t}$ is the branch current. $\Delta V_t$ represents the voltage drop vector across all branches and phases. A is a 0-1 matrix representing the three-phase topology of the distribution network. B is the matrix of the complex impedance of the line. $V_{slac}$ represents the per-unit value of the three-phase voltage at the slack bus. $\Delta V_{OLTC}$ represents the variation in voltage caused by each tap operation of the on-load tap changer (OLTC). $\rho_t$ is the integer value of the OLTC position, $\rho_{min}$ and $\rho_{max}$ represent the minimum and maximum tap positions of the OLTC transformer, respectively. The OLTC transformer is a two-winding, three-phase, Yyn0-connected transformer model [23][24].

Voltage magnitude differences in low-voltage (LV) grids play a crucial role in determining the flows in lines/cables, while voltage angles are typically small due to the high R/X ratio. As a result, a reasonable approximation for LV grids would be to consider only the real part of the voltage drop [24]. In this case, equation (13) can be substituted by:

$$\Delta V_t \approx \text{Re}\{B \cdot I_{bran,t}\} \tag{16}$$

Assuming zero voltage angles and considering only the real voltage drop is reasonable for LV networks [25], where the voltage angles are typically below 10°. As shown in equation (16), this approximation leads to a linear voltage constraint within the OPF formulation, given by:

$$V_{min} \leq V_{slac} - \Delta V_{OLTC} \cdot \rho_t + \text{Re}\{\Delta V_t\} \leq V_{max} \tag{17}$$

The balance constraints are utilized to enhance the power quality of distribution networks by ensuring the balancing of three-phase voltages.

$$VUF_{j,t}(\%) \leq VUF_{\max} \tag{18}$$

$$VUF(\%) = \frac{|V_{neg}|}{|V_{pos}|} \times 100\% \tag{19}$$

As this constraint lacks convexity, we employ an approximation of the Voltage Unbalance Factor (VUF) through the use of negative voltage sequences [26]. We assume that the positive voltage sequence is in close proximity to 1 p. u., that is, $VUF_{j,t} \approx |V_{neg,j,t}|$, in other words, $|V_{neg,j,t}| \leq VUF_{\max}$, where VUF indicates the voltage unbalance factor.

Generator constraints are considered as

$$P_{g,j,z,t}^{\min} \leq P_{g,j,z,t} \leq P_{g,j,z,t}^{\max} \tag{20}$$

$$Q_{g,j,z,t}^{\min} \leq Q_{g,j,z,t} \leq Q_{g,j,z,t}^{\max} \tag{21}$$

where $P_{g,j,z,t}^{\min}$、$P_{g,j,z,t}^{\max}$、$Q_{g,j,z,t}^{\min}$、$Q_{g,j,z,t}^{\max}$ are respectively the upper and lower limits of active and reactive power of each node j at time t in phase z.

The controllable load constraints are listed as

$$P_{d,j,z,t} = P_{1,j,z,t} + n_{j,z,t} \cdot P_{load,j,z} \tag{22}$$

$$\sum_{t=1}^{N_{hour}} n_{j,z,t} = 0 \tag{23}$$

where $P_{1,j,z,t}$ denotes the initial load demand, $P_{load,j,z}$ denotes the movable flexible load active power, and $n_{j,z,t} \in \{-1,0,1\}$, which is an integer variable that represents the change in load when transferred from $P_{1,j,z,t}$.

The constraints of battery energy storage system (BESS) are expressed as

$$SoC_{\min} \cdot E_{capa,j,z} \leq E_{j,z,t} \leq SoC_{\max} \cdot E_{capa,j,z} \tag{24}$$

$$E_{j,z,1} = E_{star} \tag{25}$$

$$E_{j,z,t} = E_{j,z,t-1} + (\eta \cdot P_{BESS,j,z,t}^{char} - \frac{P_{BESS,j,z,t}^{disc}}{\eta}) \cdot \Delta t \tag{26}$$

$$0 \leq P_{BESS,j,z,t}^{char} \leq P_{\max} \tag{27}$$

$$0 \leq P_{BESS,j,z,t}^{disc} \leq P_{\max} \tag{28}$$

$$P^{char}_{BESS,j,z,t} + P^{disc}_{BESS,j,z,t} \leq \max(P^{char}_{BESS,j,z,t}, P^{disc}_{BESS,j,z,t}) \tag{29}$$

$$Q^2_{BESS,j,z,t} \leq S^2_{max} - \max[(P^{char}_{BESS,j,z,t})^2, (P^{disc}_{BESS,j,z,t})^2] \tag{30}$$

where $E_{capa}$ represents the installed capacity of BESS, $SoC_{max}$ and $SoC_{min}$ are fixed maximum and minimum unit limits of SOC, respectively. $E_{star}$ indicates the initial power content of the BESS. The available energy $E_{j,z,t}$ is related to the BESS efficiency $\eta$. In this paper, it is defined that the battery charging and discharging power are both positive values, and the battery cannot be charged and discharged simultaneously. Other specific rules of BESS are introduced in detail in Section 3.

## 2.2. Considering the Uncertainty Through RSO

To factor in the effects of generation uncertainty and counteract possible negative consequences on security constraints, we employ a RSO model that is based on the Wasserstein distance.

### 2.2.1. RSO model

The general expression for the RSO model is as follows:

$$\begin{aligned} \min \quad & g_0(\wp, \vartheta_0) + \max_{J(\vartheta) \in \gamma} E_P[g(\wp, \vartheta)], \\ s.t. \quad & q(\wp, \vartheta) \leq 0, \quad \forall \vartheta \in q(\varpi), \\ & g(\wp, \vartheta) \in \Lambda(K, I). \end{aligned} \tag{31}$$

where $\wp$ is the decision variable representing the uncertain quantity, and $g(\wp,\vartheta)$ and $q(\wp,\vartheta)$ are its functions. $q(\varpi)$ is the uncertain set, $\varpi$ is the adjustment range of the uncertain set, and $J(\vartheta)$ denotes the set of all probability distributions on $\mathbb{R}$. $\Lambda$ is event recourse, and the corresponding definition of event fuzzy set can be expressed in the format in reference [27].

$$\Phi = \left\{ J \in \psi_0(\mathbb{R}^2 \times [N]) \middle| \begin{array}{ll} (\tilde{z}, \tilde{n}) \sim J & \\ E_J[\tilde{z} | \tilde{n} \in \varepsilon_h] \in \mathbb{N}_h & \forall h \in [K] \\ P[\tilde{z} \in Z_n | \tilde{n} = n] = 1, & \forall n \in [N] \\ P[\tilde{n} = n] = k_n & \\ \exists k \in \psi & \end{array} \right\} \tag{32}$$

Here, $\varepsilon_h$ is an event and $Z_n$ is a convex set, $n \in [N]$, $\psi \subseteq \{k \in \mathbb{R}^n_{++} | \sum_{n \in [N]} k_n = 1\}$. The random variable $\tilde{n}$ represents a set of random scenarios with uncertain probability. The confidence set of the random variable $\tilde{z}$ may be different for different scenarios, and the expectation of $\tilde{z}$ may be different conditional on the event realization.

## 2.2.2. Fuzzy set of Wasserstein distance based on 1-norm

For any $k \in [1,\infty)$, the definition of the K-type Wasserstein measure between $M$ and $M_l$ is as follows:

$$D_k(M,M_1) = \left( \inf_{\pi \in \Pi(M,M_1)} \left\{ \int \|\tau_1 - \tau_2\|^k \Pi(d\tau_1, d\tau_2) \right\} \right)^{(1/k)} \quad (33)$$

where $D_k(M,M_l)$ is the distance between $M$ and $M_l$ expressed by Wasserstein's method. $\Pi$ enables the joint probability distribution of the random variables. $M$ and $M$ are the marginal distributions of $\tau_1$ and $\tau_2$.

Considering a data-driven setting with an empirical distribution-centered Wasserstein fuzzy set $J_1 = (1/n) \sum_{n \in [N]} \rho_{l_n}$ [28], give a controllable distance metric $\psi: \mathbb{R}^{U_j} \times \mathbb{R}^{U_j} \mapsto [0,\infty)$; the Wasserstein metric defining the K-norm between two distributions $J$ and $J_1$ using an optimization problem is expressed as follows:

$$d_\tau^k(J, J_1) \triangleq \begin{cases} \inf_{\mathbb{N} \in M(J,J_1)} E_\mathbb{N} \left[ \psi(\tilde{l}, \tilde{l}^+) \right] & \text{if } k \in [1,\infty) \\ \inf_{\mathbb{N} \in M(J,J_1)} \mathbb{N} - ess \sup \psi(\tilde{l}, \tilde{l}^+) & \text{if } k=\infty \end{cases} \quad (34)$$

where $\tilde{l} \sim J$, $\tilde{l}^+ \sim J_1$ and $M(J,J_1)$ is the joint probability of the marginal distributions $J$ and $J_1$, in addition, the ambiguity set that is based on the Wasserstein distance can be represented in the following manner:

$$\mathfrak{I}_\tau^k(\zeta) = \left\{ J \in \psi_0(L) \middle| \begin{array}{l} \tilde{l} \sim J \\ d_\tau(J,J_1) \leq \zeta^k \end{array} \right\} \quad (35)$$

where $\zeta$ is the Wasserstein radius constant, $\zeta = \lambda U$, $\lambda$ is the Wasserstein radius coefficient, $U$ is an upper bound on the random demand. Thus, a new representation of Wasserstein fuzzy sets with 1-norm can be provided in the form of event fuzzy sets as follows:

$$\Phi^1 = \left\{ J \in \psi_0(\mathbb{R}^2 \times [N]) \middle| \begin{array}{l} (\tilde{z}, \tilde{n}) = ((\tilde{l}, \tilde{\sigma}), \tilde{n}) \sim J \\ E_J[\tilde{\sigma} | \tilde{n} \in [N]] \leq \zeta \\ P\left[ \begin{array}{l} \tilde{l} \in [0,L], \\ \rho(\tilde{l}, \tilde{l}_{\tilde{n}}) \leq \tilde{\sigma} \end{array} \middle| \tilde{n} = n \right] = 1, \quad \forall n \in [N] \\ P[\tilde{n} = n] = 1/N \quad \forall n \in [N] \end{array} \right\} \quad (36)$$

For practical purposes, these sets are restricted to manageable sets of conic representations, such as polyhedral or sets of second-order conic representations.

## 2.3. Overview of the RSO-AC-OPF model

This section summarizes the model structure of the RSO-OPF control method proposed in this paper for the three-phase four-wire system. Constraints (6)-(30) represent the constraints of the four-wire optimal tide, while constraint (36) represents the collapsed RSO formula. Our low-voltage distribution network model uses the RSO-OPF method to consider uncertainty while meeting BESS, CL, OLTC, and other constraints to increase the distribution network's reliability and economic value. To account for uncertainty of rooftop PV, we construct an uncertainty fuzzy set based on the Wasserstein distance of one parametric number. This fuzzy set is then affined to the event fuzzy set of the RSO model to solve the problem of PV output.

Obj: (1)

s.t: (6) - (15)

(18) - (30)

(36)

## 3. Methodology

### 3.1. Solution Process

Figure 2 illustrates the structure diagram of the uncertainty optimization control method proposed for the low-voltage distribution system. Data is preprocessed into CNN (Convolutional Neural Network) input format, followed by the application of the RSO method to remove uncertainty. Subsequently, control training is performed for each part with the CNN outputting real-time responses of the DERs, including BESS and controllable loads (CL). Cross-validation is then conducted to adjust hyperparameters to improve the accuracy of CNN classification results. Finally, the real-time response of each part is obtained. The essence of the CNN learning process is to learn the data distribution. If the distributions of the training data and the test data are different, the network's generalization ability is greatly reduced. On the other hand, if the distribution of each batch of training data is different (batch gradient), the network has to learn to adapt to various distributions in each iteration, which greatly reduces the training speed of the network. The above can be avoided by performing data preprocessing.

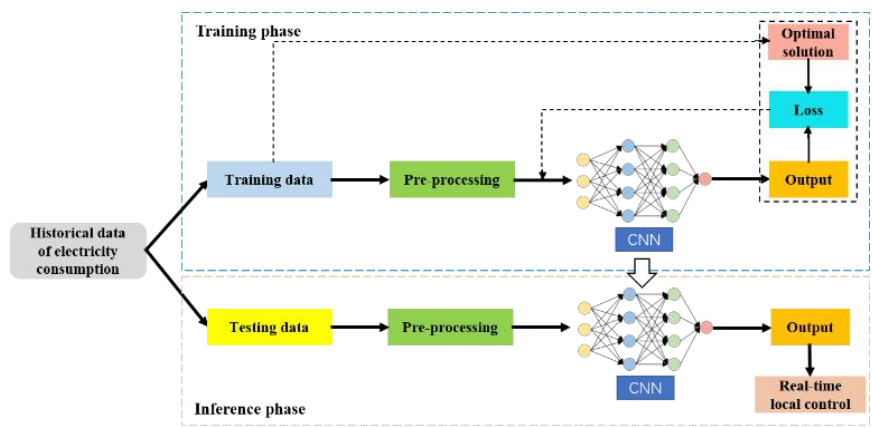

Figure 2: Frame diagram of proposed method.

## 3.2. Considering the CL through data-driven

To accommodate variable behaviours such as controllable load increase/decrease, we use a CNN model as a classifier. The model results are classified into three situations: load increase, constant, and decrease. Our CNN model uses local connection and shared weight to effectively extract potential data features. The CNN model in this paper contains a data input layer, a convolutional computation layer, a pooling layer, and a fully connected layer (INPUT-CONV-POOL-FC). The model also includes a feature extractor consisting of a convolutional layer and a sub-pooling layer, which greatly simplifies the model complexity and reduces parameters. Please refer to Algorithm 1 for details on CL real-time control and CNN model parameters. In the model of this paper, the daily load remains unchanged as a whole, and constraint (23) ensures that the categories of "load increase" and "load decrease" obtained are approximately balanced.

---
**Algorithm 1: Training CNN**

---
The data is preprocessed and labeled for each CL unit form to generate data sets. The data is divided into 80% training and 20% test sets.

**Input:** Assume a variable $\Omega=[V, P_{g,PV}]$

**Convolutional layer:** Step length: 1; the activation function is ReLU; dropout is used in the convolutional layer to prevent overfitting.

**Max Pooling layer:** The features extracted from the convolutional layer are subsampled to achieve feature reduction.

**Fully Connected layer:** The role of classification.

**Output:** The real-time response of CL (i.e., load increase or decrease), and the activation function was normalized exponential function (Softmax).

**Return:** The trained CNN model

---

## 3.3. Considering the BESS through data-driven

The BESS can absorb active power, providing an effective voltage-peaking solution. Algorithm 2 outlines the process of charging the Battery Energy Storage System (BESS) when the local PV generation exceeds the initial local load, which augments the final active power demand with the charging power. Conversely, when the local load is greater than the generating capacity, the BESS begins to discharge and reduce due to the discharge power P. The BESS's available energy is defined by the installed capacity of the BESS and the state of the battery charge, with fixed minimum and maximum unit limits. The optimization problem is similar to the CL case with $\Omega=[V, P_{PV,t}, P_{1,t}]$ as the classifier.

---
**Algorithm 2: Proposed BESS Control Design**

---
**Input:** Assume a variable $\Omega=[V, P_{PV,t}, P_{1,t}]$
**For each BESS unit form,**
**if** $P_{PV,t} \geq P_{1,t}$ ,

   **if** $(P_{PV,t} - P_{1,t}) \cdot \eta \cdot \Delta t \geq (SoC_{max} \cdot E_{capa} - E_{SoC,t-1})$

     **then** $P_{BESS,t}^{char} = \dfrac{(SoC_{max} \cdot E_{capa} - E_{SoC,t-1})}{\Delta t}$ ;

       $E_{SoC,t} = E_{SoC,t-1} + P_{BESS,t}^{char} \cdot \Delta t$

       $P_{2,t} = P_{1,t} + P_{BESS,t}^{char}$

---

```
        else  P_{BESS,t}^{char} = (P_{PV,t} - P_{1,t}) \cdot \eta
              E_{SoC,t} = E_{SoC,t-1} + P_{BESS,t}^{char} \cdot \Delta t
              P_{2,t} = P_{1,t} + P_{BESS,t}^{char}
    end if
    else if  \frac{(P_{1,t} - P_{PV,t})}{\eta} \cdot \Delta t \geq (E_{SoC,t-1} - SoC_{max} \cdot E_{capa})
    then     P_{BESS,t}^{disc} = \frac{(E_{SoC,t-1} - SoC_{min} \cdot E_{capa})}{\Delta t}
              E_{SoC,t} = E_{SoC,t-1} - P_{BESS,t}^{disc} \cdot \Delta t
              P_{2,t} = P_{1,t} - P_{BESS,t}^{disc}
        else  P_{BESS,t}^{disc} = \frac{(P_{1,t} - P_{PV,t})}{\eta}
              E_{SoC,t} = E_{SoC,t-1} - P_{BESS,t}^{disc} \cdot \Delta t
              P_{2,t} = P_{1,t} - P_{BESS,t}^{disc}
    end if
end if
Output: Real-time response of BESS
```

## 4. Simulation

The control method proposed in this paper is simulated in a modified 21-bus node low-voltage distribution network with a three-phase four-wire line, as shown in Figure 3. The rated voltage for the distribution system is 380 V. Each line section has a length of 50 m. The line self-impedance $Z_{ii}$=0.65+j0.412Ω/km and the mutual impedance $Z_{ij} = 0.01Z_{ii}$. The neutral point is assumed to be grounded at multiple points, and the capacitance is neglected due to the short cable length. The transformer grounding impedance is 3 Ω. To simulate unbalanced conditions, loads and rooftop PV panels have been unevenly distributed across the three phases. Rooftop PV icons are used to indicate PV access locations, and each bus can be single-phase or three-phase with PV installed and independently adjustable for all three phases. In addition, energy storage devices are connected at buses 4 and 16 with three phases and three phases independently adjustable and flexible loads connected to phase C of node 10, whose total daily energy consumption remains constant. $E_{capa,j,z} = 20kW \cdot h$, $\eta = 93\%$.

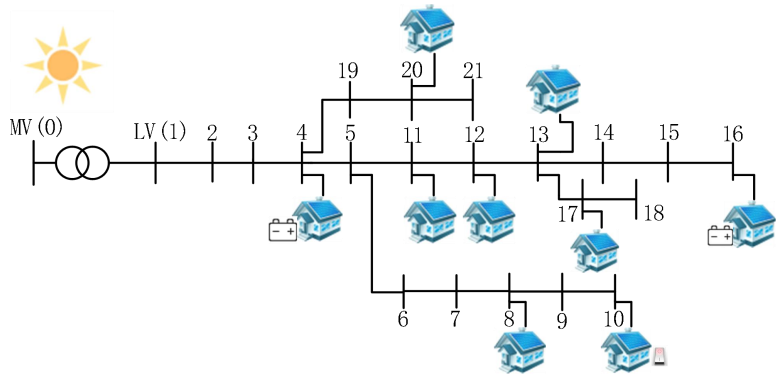

Figure 3. 21-bus low-voltage distribution network topology.

In this paper, we used 30 days of historical electricity consumption data from a residential area in the summer, with a 5-minute time resolution of photovoltaic power generation. The dataset comprises 8640 samples (30*24*12). The objective function is described in Section 2, assuming that the operating cost is $C_P = 0.2 \frac{CHF}{kWh}$, $C_Q = 0.01 \cdot C_P$.

The implementation was done in PYTHON. The optimization toolbox of RSOME [29] was used, Gurobi [30] was used as the solver, and Tensorflow was used as the platform for building the neural network. The results were obtained on HUAWEI Mate Station B515(RAM 16G, CPU AMD Ryzen 7 4700G with Radeon Graphics 3.60GHz).

### 4.1. Performance of the proposed method

We performed cross-validation, a resampling procedure, to test the performance of our controllable load model on unknown data. Figure 4 illustrates the classification accuracy during the optimization of the CNN model, with red areas indicating correct classification and blue indicating misclassification. Our training and validation resulted in obtaining a classifier with an overall accuracy of 100% due to the complete linear separability of the data. Table 1 shows that the voltage amplitude increases with PV injection. PV injection above 0.3 p. u. can cause the voltage to drop, and increasing the load with other available measures may not be sufficient to offset the reduction.

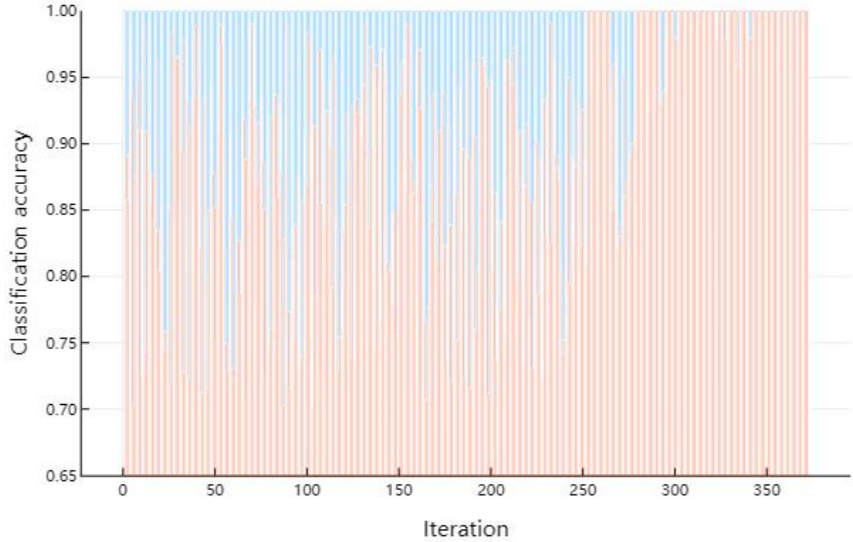

Figure 4. CNN classification accuracy.

Table 1: Real-time response results for CL.

| ID | Rooftop PV Injection [p. u.] | Voltage [p. u.] | Controllable Loads |
|---|---|---|---|
| 1 | 0 | 0.995 | increase |
| 2 | 0 | 0.996 | increase |
| 3 | 0 | 1.020 | unchanged |
| 4 | 0 | 1.017 | unchanged |
| 5 | 0 | 1.045 | decrease |
| 6 | 0 | 1.044 | decrease |
| 7 | 0.1 | 1.007 | increase |
| 8 | 0.1 | 1.0325 | unchanged |
| 9  | 0.1   | 1.032  | unchanged |
| 10 | 0.1   | 1.0575 | decrease  |
| 11 | 0.13  | 1.060  | decrease  |
| 12 | 0.2   | 1.016  | increase  |
| 13 | 0.2   | 1.043  | unchanged |
| 14 | 0.3   | 1.030  | increase  |
| 15 | 0.36  | 1.035  | increase  |
| 16 | 0.38  | 1.037  | increase  |
| 17 | 0.384 | 1.037  | increase  |
| 18 | 0.5   | 1.042  | increase  |
| 19 | 0.71  | 1.054  | increase  |
| 20 | 0.8   | 1.048  | increase  |

Figure 5 presents a visualization of real-time control for the BESS and the PV unit. The control pertains specifically to phase C of node 16. Due to the overvoltage problem, the rooftop PV absorbs reactive power to bring down the voltage and mitigate the voltage across the upper limit. When the PV power is large and cannot be entirely consumed by the grid, the BESS absorbs the excess power. For the charge state of BESS, a positive slope indicates energy storage charging, a negative slope indicates energy storage discharging; and a slope of 0 indicates no charging or discharging.

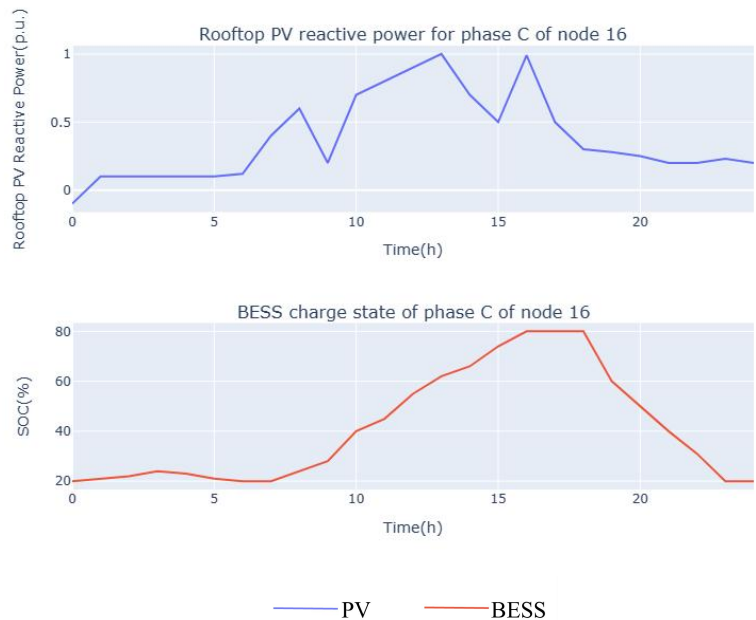

Figure 5. Real-time control of C-phase BESS and PV units at node 16.

## 4.2. Influence of the proposed method on node voltage

In Figure 6, the C-phase voltage at node 16 is compared between the control method proposed in this paper and the conventional method over a period of ten days. Node 16 is located at the end of the line and is most susceptible to voltage overruns. Due to the high power output of photovoltaic generation during the day, the upper limit of the C-phase voltage is particularly noticeable at noon. The maximum value reaches 1.09 (the standard value), but the residents are in a low phase of electricity consumption, leading to phase unbalance and other phenomena. The proposed method is superior to the conventional control method as it is capable of satisfying the voltage constraint.

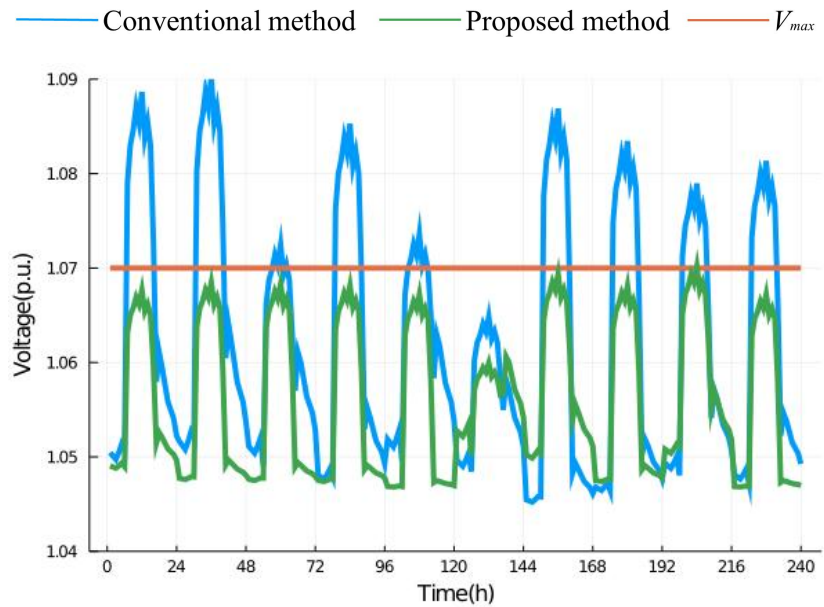

Figure 6. C-phase voltage amplitude variation at node 16 over ten days.

## 4.3. Influence of the proposed method on line current

With a maximum acceptable voltage value of 1.07 p. u. in mind, several nodes face overvoltage problems with high solar radiation during midday. Similarly, the block diagram of the thermal load of all cables shown in Figure 7 indicates that the approach of this paper allows the system cables to be free of overloads for the considered 24-hour period. Therefore, active control measures are required to keep the voltage and current below acceptable thresholds for safe grid operation.

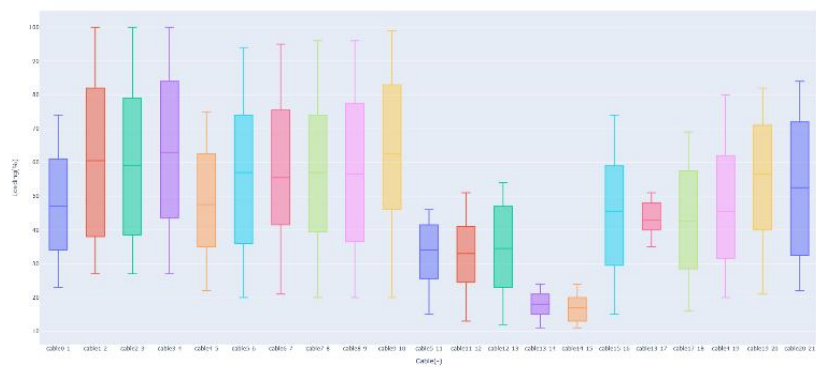

Figure 7. Box line plot of the daily current load. The box line plot displays the minimum, first quartile, median, third quartile, and maximum values for the relevant time period.

Figure 8 compares the C-phase current amplitude variation between nodes 1-2. The results indicate that the method proposed in this paper effectively suppresses the current amplitude at and below the maximum value, which ensures network security to a certain extent.

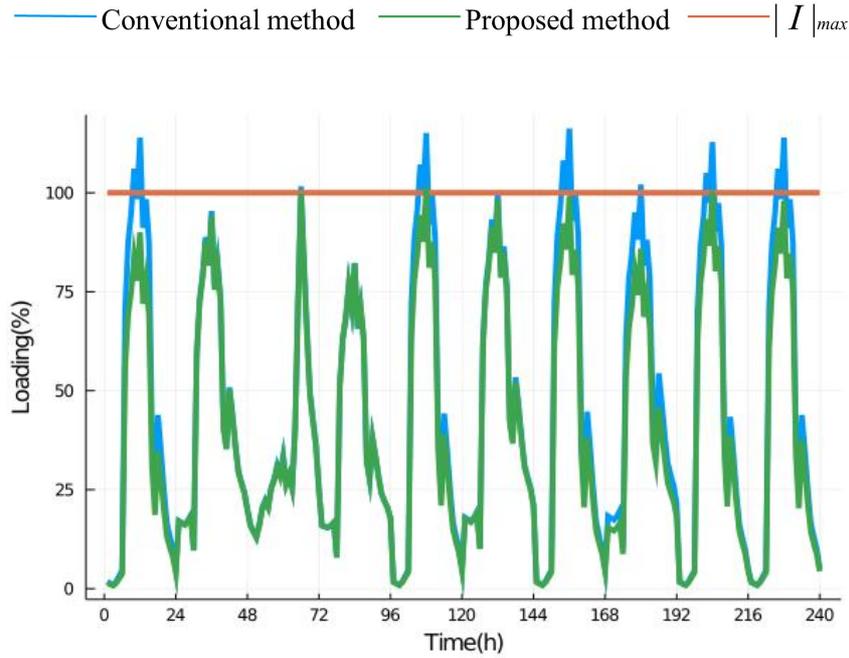

Figure 8. Variation of C-phase current amplitude between nodes 1-2.

## 4.4. Control cost and network loss comparison

The traditional distribution scheme experiences the highest total network loss when the voltage exceeds the upper limit due to the excess network loss generated by reverse power flow during the day when PV power is high. Heavier loads during the evening also increase network loss, and three-phase imbalance from PV asymmetric access results in the generation of current in the neutral line, leading to further network loss. The ideal solution assumes that all constraints are satisfied, minimizes the objective function, and has a perfect communication system. Conversely, the proposed control method in this paper controls overvoltage and overload without requiring communication facilities and improves three-phase unbalance, effectively reducing network losses and increasing economic efficiency. Table 2 summarizes the application results observed over 30 days. The validation was performed using the OpenDSS simulation platform, and the results are shown in Table 3. We confirm that the node voltages obtained from our proposed method are the same as the original results, thus further proving its validity and correctness.

Table 2: Summary of the comparison of the results under different conditions in 30 days.

| Control Solutions | $|V|_{max}$ (p. u.) | $|I|_{max}$ (%) | Losses (%) | VUF max (%) |
|---|---|---|---|---|
| Traditional | 1.09 | 117.78 | 5.89 | 1.97 |
| Ideal | 1.07 | 100 | 5 | 2 |
| This paper | 1.073 | 99.691 | 5.48 | 1.92 |

Table3: Node voltage results for both platforms

| Node number | OpenDSS | PYTHON |
|---|---|---|
| | Voltage(p. u.) | |
| 1 | 1.01 | 1.01 |
| 2 | 1.006 | 1.006 |
| 3 | 1 | 1 |
| 4 | 0.997 | 0.997 |
| 5  | 0.995 | 0.995 |
| 6  | 0.993 | 0.993 |
| 7  | 0.989 | 0.989 |
| 8  | 0.996 | 0.996 |
| 9  | 0.995 | 0.995 |
| 10 | 1.045 | 1.045 |
| 11 | 0.994 | 0.994 |
| 12 | 1.002 | 1.002 |
| 13 | 1.011 | 1.011 |
| 14 | 1.015 | 1.015 |
| 15 | 1.02  | 1.02  |
| 16 | 1.047 | 1.047 |
| 17 | 1.036 | 1.036 |
| 18 | 1.030 | 1.030 |
| 19 | 0.994 | 0.994 |
| 20 | 0.993 | 0.993 |
| 21 | 0.991 | 0.991 |

## 5. Conclusion

As the distribution system evolves, active control will play an increasingly important role in maintaining grid security, reliability, and economic efficiency. Although centralized OPF-based controllers offer optimal operation, they depend on costly monitoring and communication infrastructure. On the other hand, conventional local controllers are affordable but may struggle to adapt to rapidly changing environments and the growing penetration of DERs.

This paper investigates the challenges posed by large-scale rooftop PV access to a three-phase four-wire low-voltage distribution network, such as overvoltage. We propose a scheduling method for optimizing distributed energy sources without requiring extended monitoring or communication infrastructure. The controller is easy to compute, understand, and implement. The simulation results demonstrate that the proposed local control method can simultaneously address safety concerns in an unbalanced and uncertain three-phase four-wire system while optimizing its operation. Future work will center on the investigation of comparison of the effect of other newer deep learning algorithms for control, on the potential challenge of the use of deep learning.

## Data Availability

The data used to support the finding of this study are included within the paper.

## Conflicts of Interest

The authors declare that they have no conflicts of interest.

# Acknowledgments

This work was supported by the National Natural Science Foundation of China under Grant 51967001.